# Coexistence and interaction of spinons and magnons in an antiferromagnet with alternating antiferromagnetic and ferromagnetic quantum spin chains

H. Zhang[1], Z. Zhao[2], D. Gautreau[3], M. Raczkowski[4], A. Saha[3], V.O. Garlea[5], H. Cao[5], T. Hong[5], H. O. Jeschke[6], Subhendra D. Mahanti[1], T. Birol[3], F. F. Assaad[4,7], X. Ke[1*]

[1]Department of Physics and Astronomy, Michigan State University, East Lansing, MI 48824

[2]State Key Laboratory of Structural Chemistry, Fujian Institute of Research on the Structure of Matter, Chinese Academy of Sciences, Fuzhou, Fujian 350002, People's
Republic of China

[3]Department of Chemical Engineering and Materials Science, University of Minnesota, MN 55455

[4]Institut für Theoretische Physik und Astrophysik, Universität Würzburg, 97074 Würzburg, Germany

[5]Neutron Scattering Division, Oak Ridge National Laboratory, Oak Ridge, Tennessee 37831, USA

[6]Research Institute for Interdisciplinary Science, Okayama University, Okayama 700-8530, Japan

[7]Würzburg-Dresden Cluster of Excellence ct.qmat, Am Hubland, D-97074 Würzburg, Germany

*Corresponding author: kexiangl@msu.edu

In conventional quasi-one-dimensional antiferromagnets with quantum spins, magnetic excitations are carried by either magnons or spinons in different energy regimes: they do not coexist independently, nor could they interact with each other. In this Letter, by combining inelastic neutron scattering, quantum Monte Carlo simulations and Random Phase Approximation calculations, we report the discovery and discuss the physics of the coexistence of magnons and spinons and their interactions in Botallackite-$Cu_2(OH)_3Br$. This is a unique quantum antiferromagnet consisting of alternating ferromagnetic and antiferromagnetic Spin-1/2 chains with weak interchain couplings. Our study presents a new paradigm where one can study the interaction between two different types of magnetic quasiparticles: magnons and spinons.

In conventional magnets with magnetic long range order (LRO), low-energy excitations are carried by spin waves, represented by massless bosons called magnons with S = 1 [1]. However, in one-dimensional (1D) antiferromagnetic quantum spin systems, quantum fluctuations destroy LRO in the ground state. Such systems cannot be described using mean-field theory such as the standard Landau-Ginzburg-Wilson theory [2]. As a result, the low-energy excitations in these systems behave quite differently from their higher-dimensional counterparts. One of the prototypical systems is the Heisenberg antiferromagnetic quantum spin-1/2 chain, where the low-energy excitations are carried by pairs of deconfined spinons [3-16]. In contrast to magnons, spinons possess fractional spin S = 1/2 which could be thought of as propagating domain walls [3,4]. On the other hand, materials hosting ferromagnetic quasi-1D spin-1/2 chains are quite rare and the magnetic quasiparticles of ferromagnetic quantum spin chains are magnons [17,18].

Importantly, interaction between different quasiparticles has been an exciting research topic. In many cases, such interactions often lead to novel electronic and magnetic phenomena. For instance, electron-phonon interaction plays an essential role in the formation of Cooper pairs in conventional superconductors [19], while magnons have been proposed as the glue for Cooper pairs in unconventional superconductors [20]. In some metallic magnets, it has been found that electron-skyrmion interactions give rise to topological Hall effect [21], which provides a new route for spintronic applications. However, up to date there is no report on the interaction between two different types of magnetic quasiparticles.

In this Letter we report our observation of the coexistence and interaction of spinons and magnons in a quasi-1D antiferromagnetic insulator $Cu_2(OH)_3Br$ using inelastic neutron scattering measurements. These two different magnetic quasiparticles arise from the peculiar orbital ordering and spin structure of $Cu_2(OH)_3Br$, which consists of nearly decoupled, alternating antiferromagnetic and ferromagnetic chains of $Cu^{2+}$ ions with spin-1/2. The antiferromagnetic chains support spinons and the ferromagnetic chains support magnons. Using both quantum Monte Carlo (QMC) simulations and Random Phase Approximation (RPA) calculations, we demonstrate evidence of magnon-spinon interactions via the weak but finite interchain couplings. To the best of our knowledge, such an interaction between two different magnetic quasiparticles has not been investigated even in theory due to the unusual nature of the spin structure. Our study thus opens up a new research arena and calls for further experimental and theoretical studies.

Figure 1(a, b) depict the crystal structure of $Cu_2(OH)_3Br$, which is indicative of quasi-two-dimensional nature with the neighboring Cu-Cu distance along the c-axis much larger than those in the ab plane. The $Cu^{2+}$ magnetic ions in the ab plane form a distorted triangular lattice with two inequivalent Cu sites: Each Cu1 site has 4 Cu-O bonds and 2 Cu-Br bonds while each Cu2 site has 5 Cu-O bonds and 1 Cu-Br bond. As will be discussed later, the differences in the local geometry (caused by the ordering of Br ions) of these two Cu sites are crucial: they determine the nature of orbital ordering (partially occupied d orbitals) of Cu1 and Cu2 and the sign of nearest-neighbor intra-chain exchange interactions between Cu moments, Cu1-Cu1 and Cu2-Cu2.

Heat capacity and magnetic susceptibility measurements (inset of Fig. 1(c)) on a single crystal sample reveal a paramagnetic-antiferromagnetic phase transition at $T_N \approx 9.0$ K, in agreement with previous reports [22,23]. The main panel of Fig. 1(c) plots the temperature dependence of neutron diffraction intensity of ordering wave vector (0.5 0 0), affirming the antiferromagnetic nature of the magnetic long-range ordered state. The magnetic structure determined by Rietveld refinement (FullProf) [24] (Fig. S1) is presented in Fig. 1(d). Along the b-axis, Cu1 spins align ferromagnetically with spins oriented nearly along the diagonal direction in the ac-plane, while Cu2 spins align antiferromagnetically with spins oriented along the a-axis. The nearest-neighbor spins of both Cu1 and Cu2 sites along the a-axis are antiparallel, as suggested by the ordering wave vector. The ordered moment for Cu1 and Cu2 sites are ~ 0.737(6) $\mu_B$ and ~ 0.612(2) $\mu_B$ respectively; both of these values are smaller than the full saturation value of 1 $\mu_B$ for spin-1/2, resulting from strong quantum fluctuation.

To investigate the nature of the spin dynamics, we performed inelastic neutron scattering measurements on co-aligned single crystals in the (H K 0) scattering plane using the HYSPEC time-of-flight spectrometer at Spallation Neutron Source [27]. Intriguingly, we find that this system shows quasi-1D nature of the exchange interactions as seen in the momentum- and energy-resolved neutron scattering intensity maps presented in Fig. 2(a-c). The nearly dispersionless behavior of the excitation spectrum along both H (Fig. 2(a)) and L (Fig. 2(b)) directions indicates weak coupling between Cu spins along both a- and c-axes. In contrast, the I(E, K) intensity map (integrated over all H and L) presented in Fig. 2(c), shows unusual excitation features with well-defined magnon dispersion and broad continuum above ~ 5 meV. These observations, combined with the refined spin structure shown in Fig. 1(d), demonstrate that this system consists of nearly-decoupled, alternating ferromagnetic and antiferromagnetic chains. To the best of our knowledge, $Cu_2(OH)_3Br$ is the only system discovered thus far to exhibit the coexistence of quasi-1D ferromagnetic and antiferromagnetic quantum spin chains.

As an initial attempt to understand the magnetic excitations of this system, we performed Linear Spin Wave (LSW) calculations using SpinW [25]. The model magnetic Hamiltonian ($H$) [26] consists of nearest neighbor Heisenberg-Ising type exchange couplings with intra-chain interactions ($J_1$ and $J_2$), interchain interaction ($J_3$, $J_4$) and Dzyaloshinskii-Moriya (DM) interaction (D) (Fig. 1d). The dominant interactions are $J_1$ (ferromagnetic), $J_2$ (antiferromagnetic) whereas $J_3$ and $J_4$ are antiferromagnetic and small. The LSW fitting spectra are shown in Fig. 2(c, d) and the fitting parameters are $J_1$ = -2.6 meV, $J_2$ = 9.9 meV, $J_3$ = 1.2 meV, $J_4$ = 0.3 meV and D = 1.0 meV. The anisotropy parameter of interchain interactions is are $\Delta_F$= 0.173 for $J_1$ and $\Delta_{AF}$=0.045 for $J_2$, and the DM term is on the interchain bonds between Cu1 and Cu2 [26]. The good agreement between the experimental data and the LSW results reassures us that this system indeed is composed of quasi-1D ferromagnetic and antiferromagnetic alternating chains. The lower-energy branches associated with ferromagnetic chains have an energy gap of ~1.2 meV at the zone center (e.g. K = 0), while the higher-energy branches associated with antiferromagnetic chains have an energy gap of ~4.2 meV at the zone center (e.g., K = -1). These spin gaps arise from anisotropic exchange interactions and finite interchain coupling and the spectral gap in the ferromagnetic branch around 3.5 meV at K = -0.5 and -1.5 arises from the DM interaction.

As discussed in the introduction, the excitations of (quasi-) 1D spin-1/2 antiferromagnets are spinons instead. As a result, one expects a broad continuum produced by pairs of spinons, which cannot be described within the framework of LSW theory [7]. Indeed, we do observe a broad continuum above 5 meV as shown in Fig. 2(c), similar to the spinon continuum feature observed in the prototypical quasi-1D antiferromagnet $KCuF_3$ [5,8]. This again affirms quasi-1D nature of $Cu^{2+}$ spins of $Cu_2(OH)_3Br$.

The measured magnetic excitations and their comparison within LSW theory raise two important questions. First, what is the underlying mechanism that leads to ferromagnetic and antiferromagnetic alternating chains in this system? Second, how do the two different types of magnetic quasiparticles interact with each other?

In order to shed light on the magnetic interactions and the resultant unique spin structure of $Cu_2(OH)_3Br$, we performed first-principles density functional theory (DFT) based calculations. The total energy calculated with different long-range ordered magnetic states is listed in Fig. S4, with the lowest energy spin configuration agreeing with the experimental observation. Using only an isotropic Heisenberg model with nearest neighbor intra- and interchain couplings, the intra-chain ($J_1$ and $J_2$) and the interchain chain ($J_3$ and $J_4$) couplings, illustrated in Fig. 1(d), were calculated. Their values are listed in Fig. S5. One can see that the intra-chain interactions indeed dominate, with $J_1$ being ferromagnetic and $J_2$ antiferromagnetic. The weaker interchain couplings $J_3$ and $J_4$ are both antiferromagnetic. The theoretical results are in qualitative agreement with the exchange parameters obtained from LSW fitting. Note that spins of neighboring Cu1 and Cu2 with antiferromagnetic $J_4$ are not energetically favorable, while neighboring spins with antiferromagnetic $J_3$ are energetically favorable. The non-zero magnetic interaction $J_4$ leads to frustration, which facilitates the decoupling of Cu1 and Cu2 chains.

To understand the nature of these exchange interactions, in Fig. 3(a) we present the ground state spin density profile. The $t_{2g}$ orbitals of $Cu^{2+}$ ions are completely filled while there is a single hole in the $e_g$ manifold, which splits due to local crystal field. The spin density shows the half-filled $e_g$ orbital, which has $(x^2-y^2)$-like character in a local coordinate axis system. Interestingly, all the Cu $e_g$ orbital lobes extend towards the oxygen p orbitals but not towards the Br ions. This can be understood by the weaker crystal field associated with Br ions, which have -1 charge as opposed to -2 for the oxygen ions. The resulting crystal field pushes the Cu $e_g$ orbital with electron clouds extending towards oxygen ions to higher energies, a characteristic of the hole occupying this orbital and spin density associated with it. The crystal field, combined with the geometry and local coordinate of these two Cu sites, leads to antiferro-orbital orientational order for Cu1 chains and ferro-orbital orientational order for Cu2 chains. Such an unusual orientational ordering of the active magnetic orbital, which can be considered as an improper orbital order imposed by the strongly asymmetric crystal field of the anions, gives rise to anion-mediated exchange interactions that are dominated by Cu-O-Cu exchange pathways, considering that only O orbitals σ-bond with the half-filled Cu $e_g$ orbitals. This is supported by nearly zero spin density on the Br ions as illustrated in Fig. 3(a), which indicates that Br does not hybridize with the spin-polarized Cu orbitals, and hence does not contribute to superexchange. The projected density of states (DOS) of Br, O and the hole (i.e. the unoccupied states) of $Cu^{2+}$ ions are shown in Fig. 3(b). Consequently, antiferro-orbital order along Cu1 chains leads to ferromagnetic spin coupling ($J_1 < 0$) whereas ferro-orbital order leads to antiferromagnetic spin coupling along the Cu2 chains ($J_2 > 0$) [28].

Next, we discuss magnon-spinon interaction via the weak interchain couplings ($J_3$, $J_4$) between neighboring AFM/FM chains. In the absence of interchain couplings, the system would host deconfined spinons propagating in the AFM chain and well-defined magnons propagating in the FM chain. With gradual increase of interchain couplings, the quasi-1D nature of the system is progressively destroyed and magnetic long-range order develops. It is known that in quasi-1D antiferromagnets composed of identical spin chains, such as KCuF$_3$ [12], there is an energy threshold which separates spinons and magnons. Above this threshold, spinons are deconfined; below this threshold, the spinon continuum turns into classical magnons because of the finite interchain couplings and resulting in long-range order [13,29]. Thus, in these systems, magnetic excitations are carried either by unbound spinons or classical magnons in different energy regimes, and they do not interact. In contrast, due to the coexistence of both ferromagnetic and antiferromagnetic chains in Cu$_2$(OH)$_3$Br, the corresponding magnon and spinon excitations can coexist in the same energy range and interact with each other through the finite interchain couplings.

To better understand the effects of interchain couplings, we have used the Algorithms for Lattice Fermions (ALF) implementation [30] of the finite temperature auxiliary field quantum Monte Carlo to carry out numerical simulations of the dynamical spin structure factor of a system consisting of ferromagnetic and antiferromagnetic spin-1/2 chains [26, 31,32]. While this algorithm is formulated for fermionic systems, it can also be used to simulate non-frustrated spin systems [31]. For simplicity, we only consider intra-chain couplings ($J_1$ = -1.6 meV, $J_2$ = 5.3 meV) and antiferromagnetic interchain coupling $J_3$ while keeping $J_4$ = 0 (non-zero $J_4$ would introduce magnetic frustration and a negative sign problem).

Figure 4 presents the simulated spectra without taking into account the magnetic form factor of $Cu^{2+}$. There are several important features to point out. First, both well-defined magnon dispersion and spinon continuum, which are associated with ferromagnetic chains and antiferromagnetic chains respectively, are clearly seen, consistent with the experimental observation shown in Fig. 2(c). Second, by introducing non-zero $J_3$, the magnetic excitations associated with antiferromagnetic chains are pushed up to higher energy and a gap opens which increases with $J_3$. This gap opening is the result of molecular field arising from the neighboring ferromagnetic chains. Third, compared to the decoupled spin chains, non-zero $J_3$ introduces asymmetric spectral intensity centered about K = 1, as shown by the constant energy cut (at E = [7.7 9.7] meV) presented in Fig. 4(d), which suggests that the interchain coupling induces redistribution of spectral weight.

To obtain further insights on the effects of interchain couplings and the resultant magnon-spinon interactions, we perform Random Phase Approximation (RPA) calculations and compare the results with the INS excitation spectra. For this purpose, we adopt and generalize the RPA approach for coupled antiferromagnetic chains [33]. In the presence of interchain interaction, we obtain generalized susceptibilities $\chi_{RPA}^{F,AF}(\vec{k},\omega)$ for the two types of chains.

$$\chi_{RPA}^{F,AF}(\vec{k},\omega) = \frac{[1-J_\perp(\vec{k})\cdot\chi_{1D}^{AF,F}(k_\parallel,\omega)]\cdot\chi_{1D}^{F,AF}(k_\parallel,\omega)}{1-[J_\perp(\vec{k})]^2\cdot\chi_{1D}^{AF}(k_\parallel,\omega)\cdot\chi_{1D}^{F}(k_\parallel,\omega)}$$

(1)

$$J_\perp(\vec{k}) = 4(J_4 + J_3) \cos\left(\frac{k_\perp a}{2}\right) \cos\left(\frac{k_\parallel b}{4}\right) \qquad (2)$$

where $\chi_{1D}^{F,AF}(k_\parallel, \omega)$ are the susceptibilities of non-interacting chains and $J_\perp(k)$ is the Fourier transforms of the interchain couplings. Here $k_\parallel$ is the component of the wave vector $\vec{k}$ along the chain direction (b-axis), and $k_\perp$ is perpendicular to the chain direction (a-axis). We use a Lorentzian function for $\chi_{1D}^{F}(k_\parallel, \omega)$ and the Müller Ansatz [7] expression for $\chi_{1D}^{AF}(k_\parallel, \omega)$. Detailed description of the generalized RPA approach is documented in the Supplemental Materials [26].

Figure 5 (a, b) present the measured excitations with H integrated over [0.85 1.15] and the corresponding RPA results, respectively. In addition to the two-spinon continuum that is clearly observed in RPA calculations (Fig. 5(b)), which is consistent with the experimental data shown in Fig. 5(a), one can see a clear modification of the spectral intensity caused by the interchain couplings. For instance, a constant energy cut at E = 10.75 meV is plotted in Fig. 5(c), together with the RPA calculations with (red) and without (black) interchain couplings. One can see that RPA with the inclusion of interchain couplings captures the redistribution of the spectral weight with the intensity at K = -0.5 larger than that at K = -1.5. This difference cannot be accounted for by magnetic form factor. Note that $J_\perp(\vec{k})$ (Eq. (2)) is negative when K is in the range of [-1 0] and positive when K is in the range [-2 -1]. This difference in the sign leads to the asymmetry in the spectral weight about K = -1, which is consistent with the QMC simulation results shown in Fig. 4(b-d). If we reduce the constant energy cut to E = 7.75 meV (Fig. 5(d)) and focus on the two peaks closest to K

= -1, again the RPA spectrum with interchain couplings introduces asymmetry. The agreement near K = -1.25 is very good but not so good for K = -0.75. Further comparison between experimental data and RPA calculation results are discussed in the Supplemental Materials [26].

In summary, we have discovered that magnons and spinons coexist in $Cu_2(OH)_3Br$, which uniquely consists of quasi-1D ferromagnetic and antiferromagnetic quantum spin chains. Magnons and spinons interact with each other via weak but finite interchain couplings, which opens the gap of the spinon continuum and gives rise to a redistribution of the spectral weight. This study highlights a new toy model and research paradigm to study the interaction between two different types of magnetic quasiparticles.

**Acknowledgement**

X. K. acknowledges the financial support by the U.S. Department of Energy, Office of Science, Office of Basic Energy Sciences, Materials Sciences and Engineering Division under DE-SC0019259, and is grateful to Prof. G.-W Chern and Prof. A. L. Chernyshev for insightful discussion. H. Z. is supported by the National Science Foundation under DMR-1608752. T. B. and D. G. are funded by the Department of Energy through the University of Minnesota Center for Quantum Materials under DE-SC-0016371 and acknowledge the Minnesota Supercomputing Institute (MSI) at the University of Minnesota for providing resources that contributed to the research results reported within this paper. A portion of this research used resources at both High Flux Isotope Reactor and Spallation Neutron Source, which are DOE Office of Science User

Facilities operated by the Oak Ridge National Laboratory. Z. Z. acknowledges the supports from the National Natural Science Foundation of China (Grants NO. U1832166 and No. 51702320). The authors gratefully acknowledge the Gauss Centre for Supercomputing e.V. (www.gauss-centre.eu) for funding this project by providing computing time on the GCS Supercomputer SUPERMUC-NG at Leibniz Supercomputing Centre (www.lrz.de) (Project-id pr53ju). F.F.A. thanks funding from the Deutsche Forschungsgemeinschaft under the grant number AS 120/14-1 for the further development of the Algorithms for Lattice Fermions QMC code, as well as through the Würzburg-Dresden Cluster of Excellence on Complexity and Topology in Quantum Matter - ct.qmat (EXC 2147, project-id 390858490). M.R. is supported by the German Research Foundation (DFG) through Grant No. RA 2990/1- 1.

**Figure Captions:**

Figure 1. **Crystal structure and magnetic structure of $Cu_2(OH)_3Br$.** Crystal structure of $Cu_2(OH)_3Br$ in the *ac* (a) and *ab* (b) plane showing a quasi-two dimensional, distorted triangular lattice of Cu atoms. (c) Temperature dependence of neutron diffraction intensity of an ordering wave vector (0.5 0 0). The inset shows the temperature dependence of heat capacity and magnetic susceptibility measurements. (d) Schematics of spin structure of $Cu^{2+}$ ions with Cu2 spins point along the a-axis while Cu1 spins pointing nearly along the diagonal direction in the *ac* plane. Exchange interactions of Cu1-Cu1, Cu2-Cu2, and Cu1-Cu2 as well as DM interaction are denoted.

Figure 2. **Magnetic excitation spectra and the comparison to LSW calculations.** (a) The momentum- and energy-resolved neutron scattering intensity map I(E, H) (K = -0.5 and with all measured L values integrated). (b) Intensity map I(E, L) (K = -0.5 and with all measured H values integrated). These two intensity maps show nearly dispersionless magnetic excitations along both H and L directions. (c) Intensity map I(E, K) with both H and L integrated over all measured values to enhance the statistics of the signal. These intensity maps were obtained after using the data measured at $T$ = 100 K as background and subtracting it from the data measured at $T$ = 5 K. (d) The calculated I(E, K) spectra using LSW theory. The white curves in all panels are the calculated dispersions using LSW theory.

Figure 3. **Electronic structure calculated via first principles DFT.** (a) The ground state spin density of the half-filled $e_g$ orbital of $Cu^{2+}$ ions and p orbitals of O and Br atoms. Yellow color denotes spin up and cyan color denotes spin down. Cu1 ions with ferromagnetic spin alignment show antiferro-orbital order while Cu2 ions with antiferromagnetic spin alignment show ferro-orbital order. (b) The projected density of states (PDOS) of Cu1, Cu2, Br, and O ions.

Figure 4. **Magnetic excitation spectra via quantum Monte Carlo simulations.** Simulated magnetic excitation spectra (with H = 1) of a system consisting of alternating ferromagnetic and antiferromagnetic quantum spin chains with the interchain coupling $J_3 = 0$ (a), $J_3 = 0.1J_2$ (b), and $J_3 = 0.2J_2$ (c). (d) Constant energy cuts at E = [8.7 9.7] meV showing the asymmetric spectral intensity about K = 1 induced by non-zero $J_3$. Note that Bose factor but not magnetic form factor of $Cu^{2+}$ ions has been taken into account in the simulation.

Figure 5. **Magnetic excitation spectra and the comparison with RPA calculations.** (a) I(E, K) intensity map obtained after background subtraction with H integrated over [0.85 1.1] and L integrated over all measured values. (b) The RPA calculation of I(E, K) spectra for comparison. Constant energy cuts at E = 10.75 meV (c) and at E = 7.75 meV (d) and their comparison with RPA calculations.

Figure 1.
H. Zhang et al

(a)

(b)

- Cu (black)
- Cu (blue)
- O
- Br

(c)

(d)

- MCu1
- MCu2
- J1
- J2
- J3
- J4
- DM1
- DM2

Figure 2.
H. Zhang et al,

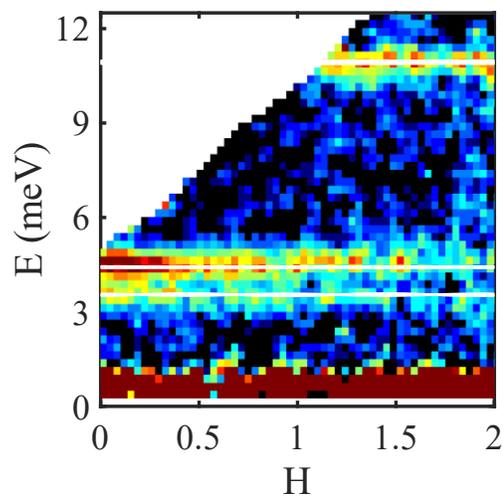
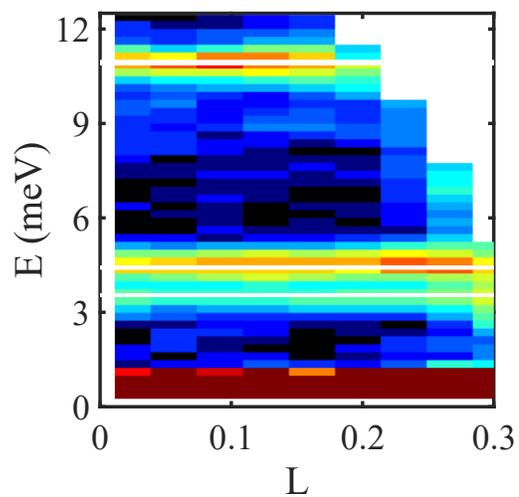
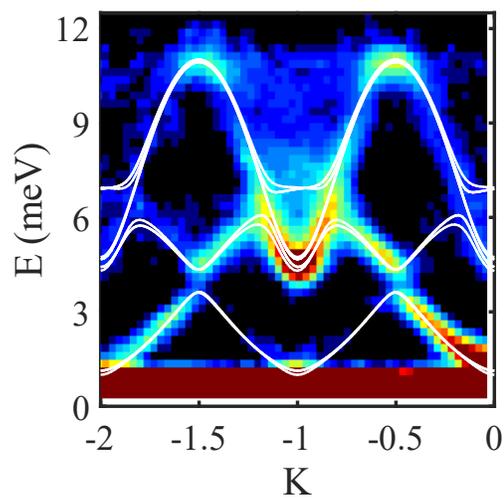
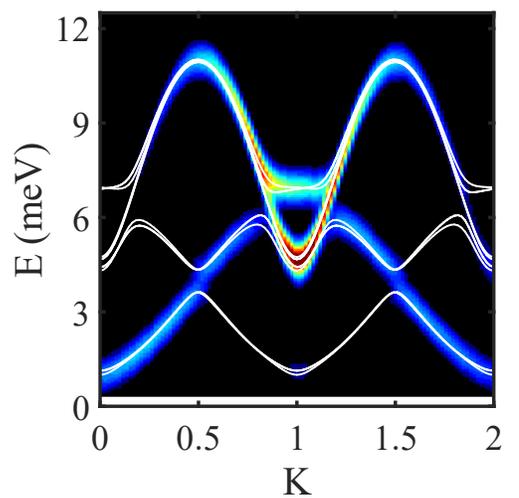

Figure 3.
H. Zhang et al,

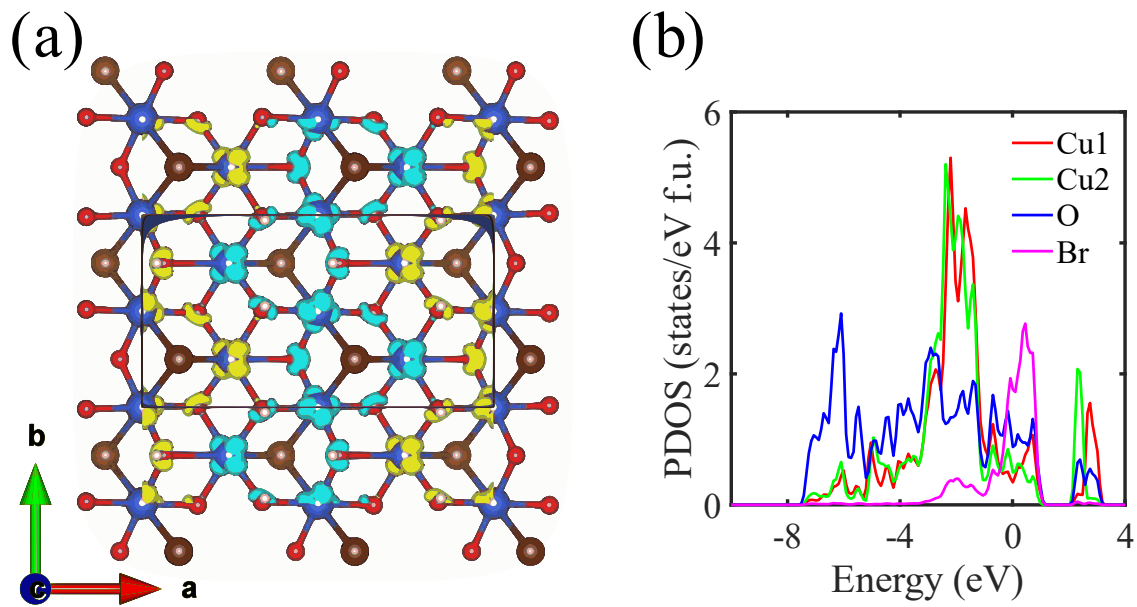

Figure 4.
H. Zhang et al,

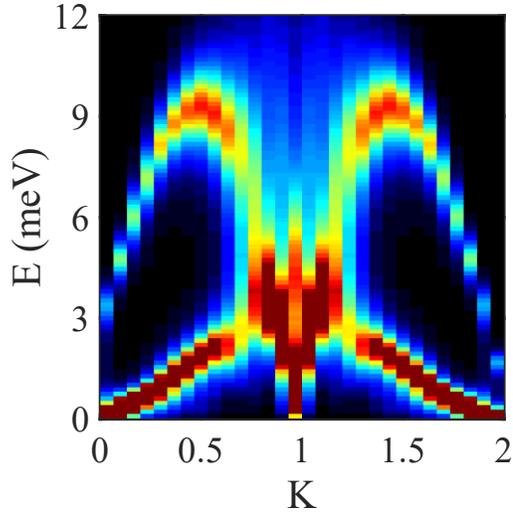
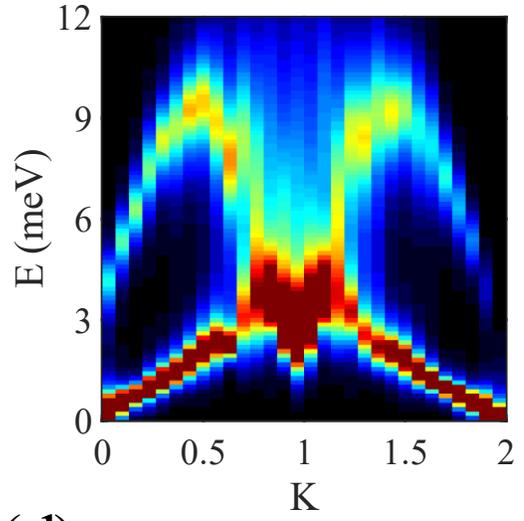
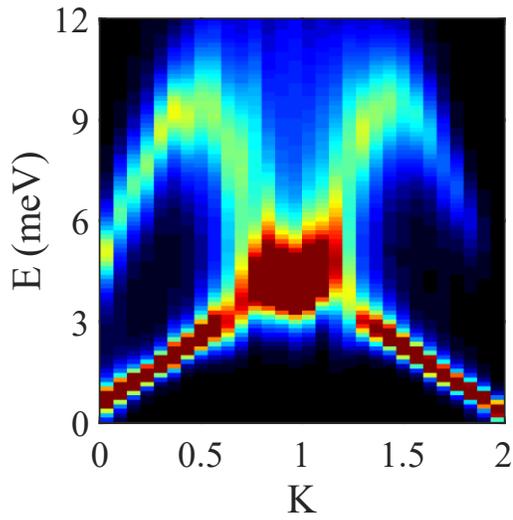
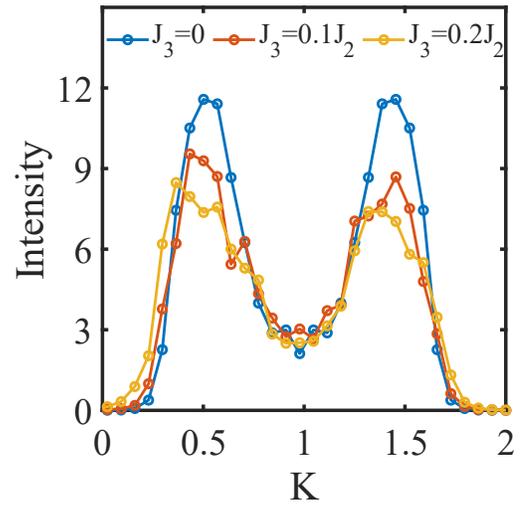

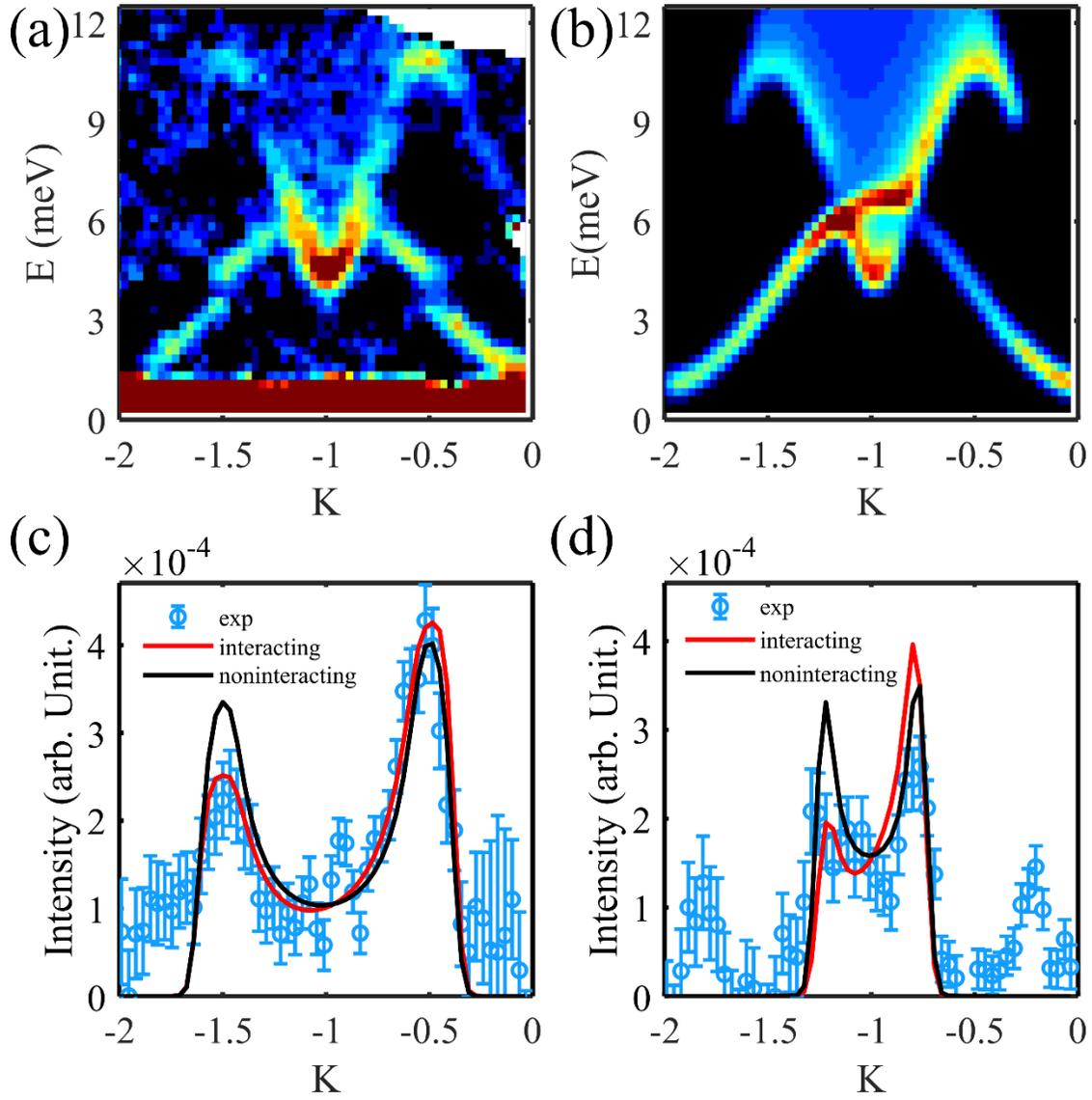

# References

[1] C. Kittel and P. McEuen, *Introduction to solid state physics* (Wiley New York, 1996), Vol. 8.
[2] L. D. Landau, E. M. Lifshitz, and E. M. Pitaevskii, (Butterworth-Heinemann, New York, 1999).
[3] L. D. Faddeev and L. A. Takhtajan, Physics Letters A **85**, 375 (1981).
[4] F. D. M. Haldane, Physical Review Letters **66**, 1529 (1991).
[5] S. E. Nagler, D. A. Tennant, R. A. Cowley, T. G. Perring, and S. K. Satija, Physical Review B **44**, 12361 (1991).
[6] D. A. Tennant, S. E. Nagler, D. Welz, G. Shirane, and K. Yamada, Physical Review B **52**, 13381 (1995).
[7] M. Karbach, G. Müller, A. H. Bougourzi, A. Fledderjohann, and K.-H. Mütter, Physical Review B **55**, 12510 (1997).
[8] B. Lake, D. A. Tennant, and S. E. Nagler, Physical Review Letters **85**, 832 (2000).
[9] A. Zheludev *et al.*, Physical Review Letters **85**, 4799 (2000).
[10] A. Zheludev, M. Kenzelmann, S. Raymond, T. Masuda, K. Uchinokura, and S. H. Lee, Physical Review B **65**, 014402 (2001).
[11] R. Coldea, D. A. Tennant, A. M. Tsvelik, and Z. Tylczynski, Physical Review Letters **86**, 1335 (2001).
[12] B. Lake, D. A. Tennant, C. D. Frost, and S. E. Nagler, Nature Materials **4**, 329 (2005).
[13] M. Raczkowski and F. F. Assaad, Physical Review B **88**, 085120 (2013).
[14] M. Mourigal, M. Enderle, A. Klöpperpieper, J.-S. Caux, A. Stunault, and H. M. Rønnow, Nature Physics **9**, 435 (2013).
[15] A. K. Bera, B. Lake, F. H. L. Essler, L. Vanderstraeten, C. Hubig, U. Schollwöck, A. T. M. N. Islam, A. Schneidewind, and D. L. Quintero-Castro, Physical Review B **96**, 054423 (2017).
[16] B. Lake, A. M. Tsvelik, S. Notbohm, D. Alan Tennant, T. G. Perring, M. Reehuis, C. Sekar, G. Krabbes, and B. Büchner, Nature Physics **6**, 50 (2010).
[17] M. Takahashi, P. Turek, Y. Nakazawa, M. Tamura, K. Nozawa, D. Shiomi, M. Ishikawa, and M. Kinoshita, Physical Review Letters **67**, 746 (1991).
[18] S. K. Satija, J. D. Axe, R. Gaura, R. Willett, and C. P. Landee, Physical Review B **25**, 6855 (1982).
[19] J. Bardeen, L. N. Cooper, and J. R. Schrieffer, Physical Review **106**, 162 (1957).
[20] D. J. Scalapino, Reviews of Modern Physics **84**, 1383 (2012).
[21] A. Neubauer, C. Pfleiderer, B. Binz, A. Rosch, R. Ritz, P. G. Niklowitz, and P. Böni, Physical Review Letters **102**, 186602 (2009).
[22] Z. Y. Zhao, H. L. Che, R. Chen, J. F. Wang, X. F. Sun, and Z. Z. He, J Phys-Condens Mat **31** (2019).
[23] X. G. Zheng, T. Yamashita, M. Hagihala, M. Fujihala, and T. Kawae, Physica B: Condensed Matter **404**, 680 (2009).
[24] J. Rodríguez-Carvajal, Physica B: Condensed Matter **192**, 55 (1993).
[25] S. Toth and B. Lake, J Phys-Condens Mat **27** (2015).

[26] See Supplemental Materials for detials on experimental methods, Rietveld refinement, linear spin wave fitting, calculations of two-magnon continuum bounds, DFT calculations, quantum Monte Carlo simulations, and Random Phase Approximation calculations, which include Refs. [34-44].
[27] B. Winn, U. Filges, V. O. Garlea, M. Graves-Brook, M. Hagen, C. Jiang, M. Kenzelmann, L. Passell, S. M. Shapiro, X. Tong et al., EPJ Web of Conferences 83, 03017 (2015).
[28] D. I. Khomskii, Physica Scripta **72**, CC8 (2005).
[29] M. Kohno, O. A. Starykh, and L. Balents, Nature Physics **3**, 790 (2007).
[30] M. Bercx, F. Goth, J. S. Hofmann, and F. F. Assaad, SciPost Phys. **3**, 013 (2017).
[31] R. Blankenbecler, D. J. Scalapino, and R. L. Sugar, Physical Review D **24**, 2278 (1981).
[32] F. Assaad and H. Evertz, *Computational Many-Particle Physics* (Springer-Verlag, Berlin, 2008), Vol. 739, Lecture Notes in Physics, p.^pp. 277-356.
[33] D. J. Scalapino, Y. Imry, and P. Pincus, Physical Review B **11**, 2042 (1975).
[34] T. Hong *et al.*, Physical Review B **74**, 094434 (2006).
[35] M. B. Stone, I. A. Zaliznyak, T. Hong, C. L. Broholm, and D. H. Reich, Nature **440**, 187 (2006).
[36] T. Huberman, R. Coldea, R. A. Cowley, D. A. Tennant, R. L. Leheny, R. J. Christianson, and C. D. Frost, Physical Review B **72**, 014413 (2005).
[37] G. Kresse and J. Furthmüller, Physical Review B **54**, 11169 (1996).
[38] S. L. Dudarev, G. A. Botton, S. Y. Savrasov, C. J. Humphreys, and A. P. Sutton, Physical Review B **57**, 1505 (1998).
[39] G. Kresse and J. Hafner, Physical Review B **47**, 558 (1993).
[40] G. Kresse and D. Joubert, Physical Review B **59**, 1758 (1999).
[41] J. P. Perdew, K. Burke, and M. Ernzerhof, Physical Review Letters **77**, 3865 (1996).
[42] K. Koepernik and H. Eschrig, Physical Review B **59**, 1743 (1999).
[43] A. W. Sandvik, Physical Review B **57**, 10287 (1998).
[44] G. Reiter and A. Sjolander, Journal of Physics C: Solid State Physics **13**, 3027 (1980).
[45] B. Winn, U. Filges, V. O. Garlea, M. Graves-Brook, M. Hagen, C. Jiang, M. Kenzelmann, L. Passell, S. M. Shapiro, X. Tong et al., EPJ Web of Conferences 83, 03017 (2015).